\documentclass[aps,pra,reprint,groupedaddress,floatfix]{revtex4-2}
\usepackage{amssymb}
\usepackage{xcolor,soul}
\usepackage{graphicx}% Include figure files
\usepackage{dcolumn}% Align table columns on decimal point
\usepackage{bm}% bold math

\usepackage[utf8]{inputenc}
\usepackage[T1]{fontenc}
\usepackage{mathptmx}

\newcommand\Be{$^{9} \text{Be}^+$}
\newcommand\Ca{$^{40} \text{Ca}^+$}
\newcommand\coherence{$1.3(4)~\text{s}$}

\begin{document}

\title[]{Second-Scale \Be~Spin Coherence in a Compact Penning Trap}

\author{Brian J. McMahon}
\email{Brian.McMahon@gtri.gatech.edu}
\author{Brian C. Sawyer}
\affiliation{Georgia Tech Research Institute, Atlanta, Georgia 30332, USA}

\date{\today}

\begin{abstract}
We report microwave spectroscopy of co-trapped \Be~and \Ca~within a compact permanent-magnet-based Penning ion trap.  The trap is constructed with a reconfigurable array of NdFeB rings providing a 0.654~T magnetic field that is near the 0.6774-T magnetic-field-insensitive hyperfine transition in \Be. Performing Ramsey spectroscopy on this hyperfine transition, we demonstrate nuclear spin coherence with a contrast decay time of $>1~\text{s}$. The \Be~is sympathetically cooled by a Coulomb crystal of \Ca, which minimizes \Be~illumination and thus mitigates reactive loss. Introducing a unique high-magnetic-field optical detection scheme for \Ca, we perform spin state readout without a 729~nm shelving laser. We record a fractional trap magnetic field instability below $20~\text{ppb}$ ($<13$~nT) at 43~s of averaging time with no magnetic shielding and only passive thermal isolation. We discuss potential applications of this compact, reconfigurable Penning trap.   
\end{abstract}

\maketitle

\section{Introduction} 

Cold, trapped atomic ions have long served as platforms for precision measurements. In contrast with neutral atoms and molecules, the ion charge provides a mechanism for stable, long-term confinement that is insensitive to internal state dynamics. Penning traps, which confine ions via a combination of static electric and magnetic fields, have facilitated early Doppler laser cooling demonstrations \cite{wineland_radiation-pressure_1978,itano_laser_1982,itano_perpendicular_1988}, studies of non-neutral plasma dynamics~\cite{bollinger_nonneutral_1994, dubin_trapped_1999}, precision tests of quantum electrodynamics \cite{kohler_isotope_2016}, laboratory-scale atomic clocks~\cite{bollinger_303-mhz_1991}, and quantum information processing/simulation~\cite{biercuk_high-fidelity_2003,crick_fast_2010,britton_engineered_2012, bohnet_quantum_2016, garttner_measuring_2017, jain_scalable_2020}. Penning traps are particularly well-suited for precision measurements of masses~\cite{bradley_penning_1999,schussler_detection_2020,rau_penning_2020} and gyromagnetic ratios~\cite{hanneke_new_2008}. 

A number of compact permanent magnet Penning traps (PMPTs) have been experimentally demonstrated~\cite{lemaire_compact_2018,gomer_compact_1995,suess_permanent_2002,brewer_lifetime_2018, mcmahon_doppler-cooled_2020}, but the magnetic field instabilities of such compact traps -- a key metric influencing the stability of atomic transition frequencies -- have yet to be quantified. Permanent magnet assemblies have some advantages over typical superconducting magnets including small size and room-temperature operation. Here we characterize the long-term magnetic field instability of a PMPT via microwave spectroscopy of a Coulomb crystal of \Ca, measuring $<20$~ppb field fluctuations at 43 s and a drift-removed field instability below 10~ppb at 2-3~min. 

\begin{figure}
    \centering
	\includegraphics[width=0.48\textwidth]{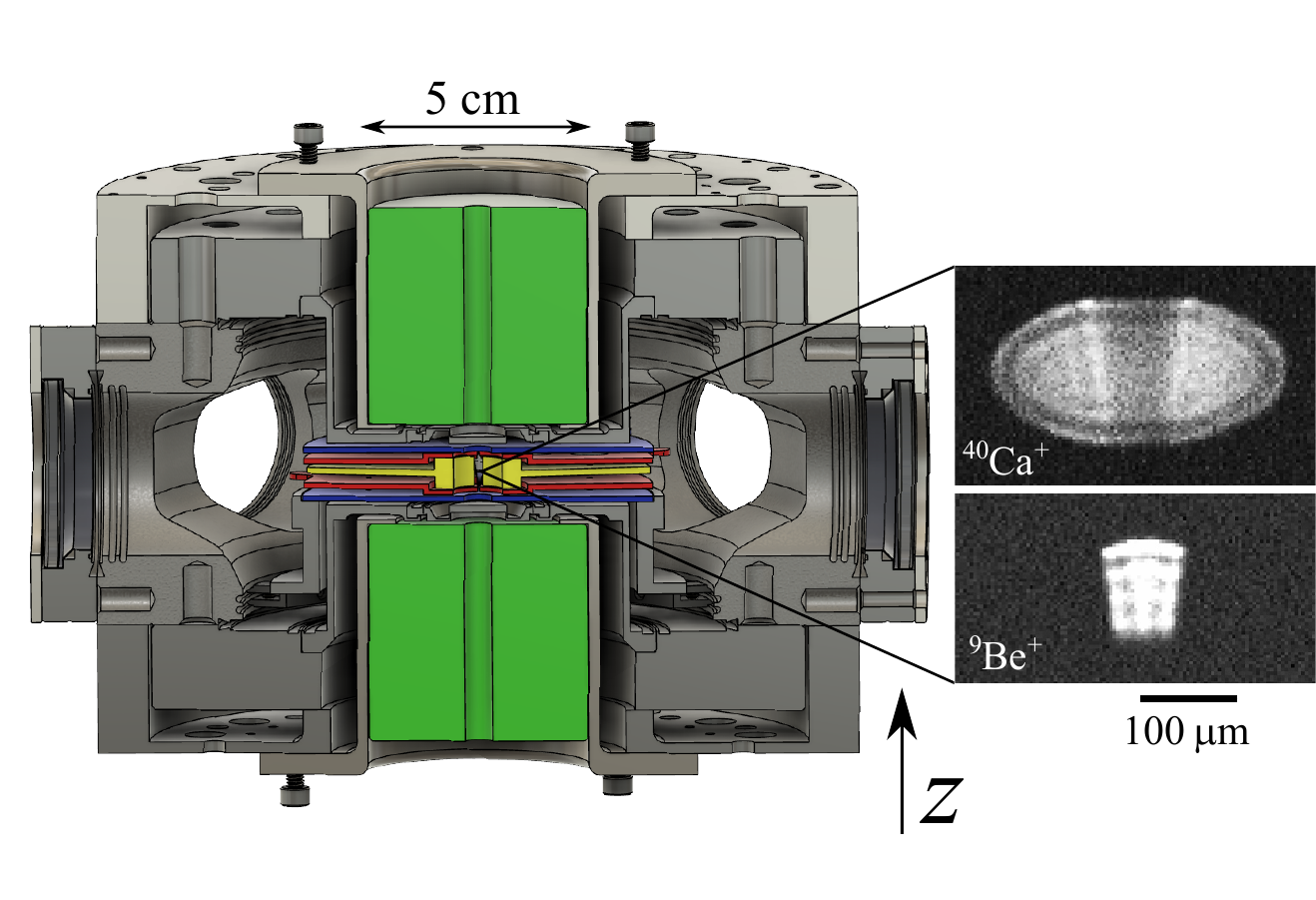}
	\caption{\label{fig:Trap} Cross section view illustration (left) of the compact Penning ion trap with N52 NdFeB permanent ring magnets (green cylinders) and side-view images of co-trapped, Doppler-cooled \Ca~(upper right image) and \Be~(lower right image). The two images are recorded separately by translating a fused silica objective to focus either the 393~nm (\Ca) or 313~nm (\Be) fluorescence on an electron multiplying charge coupled device camera. For these experiments, we trap $\sim100$ \Be~and $\sim10^3$ \Ca~as shown.}
\end{figure} 

Many atomic ions with a nuclear quadrupole moment (i.e. nuclear spin $I\geq 1$) possess ground-state `clock' transitions at non-zero magnetic field that are first-order insensitive to field fluctuations. For example, the lower-field ($<0.1~\text{T}$) clock transitions in \Be~\cite{gaebler_high-fidelity_2016} and $^{43}\text{Ca}^+$~\cite{ballance_high-fidelity_2016} have been used for high-fidelity quantum entangling gates, while higher-field clock transitions exhibit extraordinary decoupling from magnetic field noise with typical residual sensitivities of $<100~\text{Hz/mT}^2$. For \Be~near 0.6774~T (0.8194~T), the residual quadratic field sensitivity of the 321-MHz (303-MHz) clock transition is $\sim15~\text{Hz}/\text{mT}^2$. Notably, Bollinger and co-authors observed a record Ramsey free precession time~\footnote{To our knowledge, the result of Bollinger et al. (1991) is the longest reported Ramsey coherence in the absence of dynamical decoupling.} of 550~s with \Be~at $0.8194~\text{T}$ in a Penning trap atomic clock, measuring a fractional frequency instability of $1\times10^{-13}$~\cite{bollinger_303-mhz_1991}. 

In this article, we report the first Rabi and Ramsey spectroscopy measurements of \Be~using co-trapped \Ca~as a sympathetic coolant. Unlike with Be$^+$ or Mg$^+$, chemical reactions between Ca$^+$ and background H$_2$ are naturally reversed under illumination by the Ca$^+$ Doppler cooling light~\cite{condoluci_reassigning_2017}. We measure a \Be~spin coherence duration of $>1~\text{s}$ under ambient conditions by operating near the 0.6774-T magnetic-field-insensitive (MFI) transition. This demonstration introduces \Be--\Ca~mixed-species crystals for Penning trap experiments and lays the groundwork for precision spectroscopy and portable atomic clocks at high magnetic field using PMPTs. We describe the potential for PMPT \Be~atomic clock operation below $10^{-16}$ fractional frequency error, which would compare favorably with recent compact trapped-ion microwave clocks~\cite{schwindt_highly_2016,ely_using_2018,burt_demonstration_2021}.

\section{Trap Design, Doppler Cooling, and Detection}

In a previous publication, we reported Doppler laser cooling of \Ca~in a compact, reconfigurable Penning trap~\cite{mcmahon_doppler-cooled_2020}. Here, we use the same trap electrode assembly as in Ref.~\cite{mcmahon_doppler-cooled_2020}, which consists of a stack of Ti plate electrodes spaced by sapphire beads. For this work, we have added optical access along the magnetic field ($z$) axis by installing re-entrant sapphire viewports and NdFeB permanent ring magnets (see Fig.~\ref{fig:Trap}). The two identical N52 NdFeB magnets are axially magnetized with dimensions of 6.35 mm (inner diameter) $\times$ 50.8 mm (outer diameter) $\times$ 50.8 mm (length). The inner faces of the magnets are spaced by $\sim24~\text{mm}$ to minimize magnetic field curvature at the \Be--\Ca~crystal location~\cite{frerichs_analytic_1992,mcmahon_doppler-cooled_2020}. With this permanent magnet configuration, we achieve a trap magnetic field of 0.654~T with a corresponding bare cyclotron frequency ($f_c=qB/2\pi m$) of 1.11~MHz (251~kHz) for \Be~(\Ca). Here $q$ and $m$ are the charge and mass of a single ion, respectively, and $B$ is the trap magnetic field. The trap electrode voltages are chosen to create a harmonic axial confining potential with a center-of-mass frequency ($f_z$) of 232~kHz (110~kHz) for \Be~(\Ca). The modified cyclotron ($f_+$) and magnetron ($f_-$) frequencies are
\begin{equation}\label{eq:cycmag}
    f_{\pm} = \frac{f_c}{2}\left( 1 \pm \sqrt{1-2\frac{f_z^2}{f_c^2}} \right).
\end{equation}
Equation~\ref{eq:cycmag} yields $f_+ = 1.09~\text{MHz}~(224~\text{kHz})$ and $f_-= 24.7~\text{kHz}~(27.0~\text{kHz})$ for \Be~(\Ca). Simultaneous stable confinement of both species requires a global crystal rotation frequency, $f_r$, such that $27~\text{kHz} < f_r < 224~\text{kHz}$. We perform these experiments with $f_r\sim60~\text{kHz}$. The \Be--\Ca~crystal rotation is stabilized via an axialization drive~\cite{phillips_dynamics_2008} at the $f_c$ of \Ca~or a quadrupole rotating wall potential at the intended $f_r$~\cite{hasegawa_stability_2005}, with similar spectroscopic results in either case. For atomic clock operation, a rotating wall potential is ultimately preferred for minimization of the second-order-Doppler shift uncertainty~\cite{tan_minimizing_1995} via phase-locked control of $f_r$.

Figure~\ref{fig:Trap}(a) shows an illustration of the current Penning trap assembly. Neutral fluxes of Be and Ca are produced separately from in-vacuum, resistively-heated ovens placed outside the trap electrodes in the radial direction. We begin dual-species experiments by ionizing Ca with a combination of a resonant 423 nm laser beam and a non-resonant 313 nm laser beam (the same used for \Be~detection). The 313 nm beam is pulsed on for only a few seconds to control the total number of trapped \Ca. Once cold \Ca~is observed in the trap, we heat the Be oven and ionize with the fourth harmonic of a 30-ps pulsed Nd:YAG laser at 266 nm. When both ion species are loaded, we can separately image their fluorescence as shown in Figs.~\ref{fig:Trap}(b-c) using a fused silica side-view objective with a numerical aperture of 0.2 and a working distance of 117~mm (121~mm) for \Be~(\Ca).

\begin{figure}
    \centering
	\includegraphics[scale=0.31]{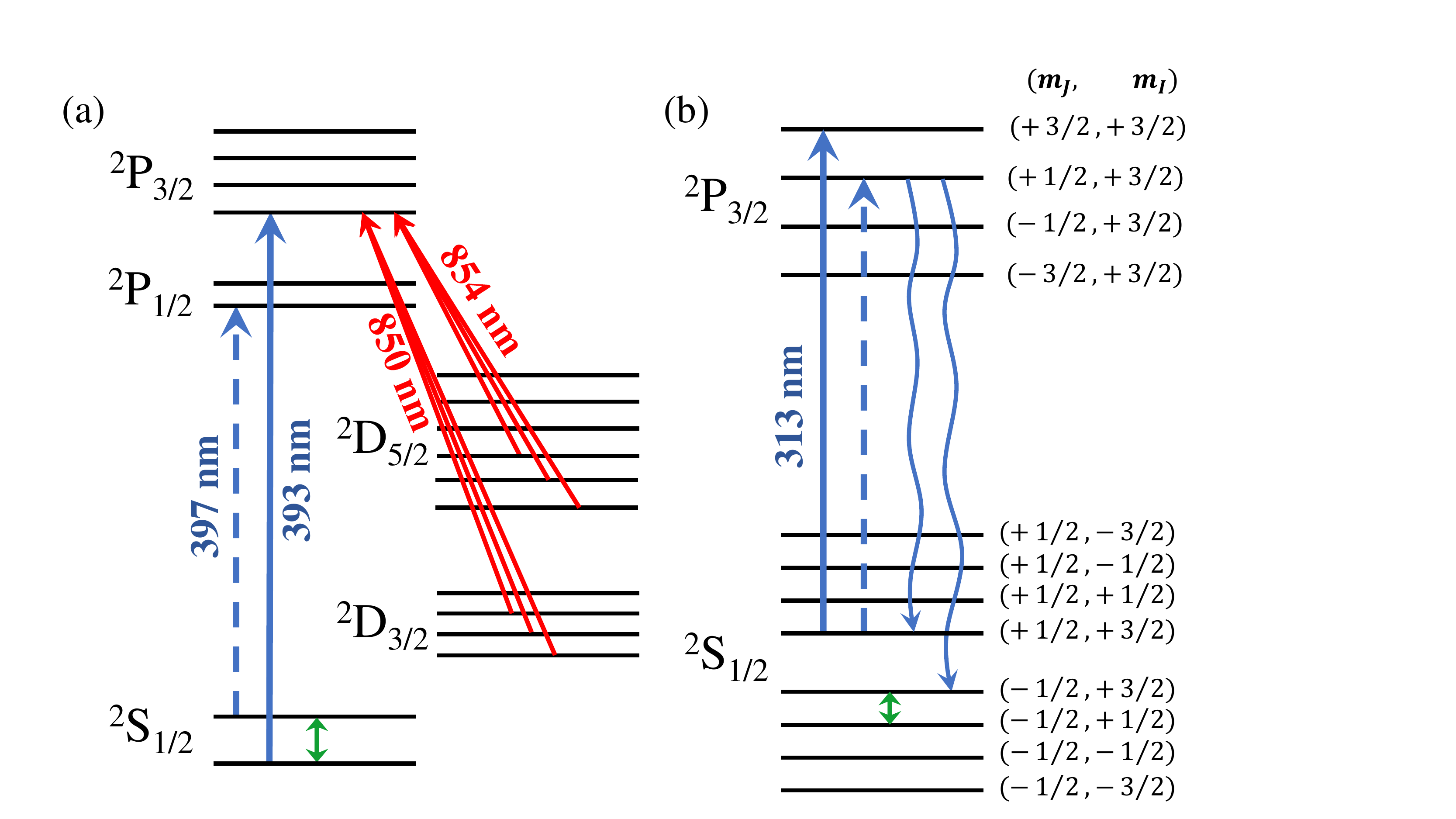}
	\caption{\label{fig:CaBe} Simplified energy level diagrams for (a) \Ca~and (b) \Be~(not to scale). The microwave and radiofrequency transitions interrogated for this work are labelled with green arrows. The dashed arrow in (a) represents the weak 397-nm repump laser, while the dashed arrow in (b) denotes the optical pumping wavelength for preparation to the upper state of the \Be~MFI transition.}
\end{figure} 

In order to differentiate between the ground state populations of \Ca~without a narrow-band 729~nm shelving laser, we use a non-standard detection scheme. As illustrated in Fig.~\ref{fig:CaBe}(a), we apply a 393 nm laser beam tuned $\sim 20$~MHz below the $|S_{1/2},m_J=-1/2\rangle \rightarrow |P_{3/2}, -3/2\rangle$ electronic transition, where $m_J$ is the projection of total electronic angular momentum on the magnetic field axis. Spontaneous decays via electric dipole transitions from the $|P_{3/2},-3/2\rangle$ state repopulate the initial $|S_{1/2},-1/2\rangle$ state as well as the $D_{3/2}$ and $D_{5/2}$ metastable levels. In order to minimize population of the `dark' $|S_{1/2},+1/2\rangle$ state, we preferentially repump from the metastable levels back to $|P_{3/2},-3/2\rangle$ using 850~nm and 854~nm laser beams. All $m_J$ sublevels are fully resolved at our trap magnetic field of $0.654~\text{T}$, therefore, we use sidebands produced in optical-fiber-based electro-optic modulators (EOMs) to clear the metastable D-states \cite{koo_doppler_2004,mcmahon_doppler-cooled_2020}. This configuration produces a quasi-closed cycle that excludes the $|S_{1/2}, +1/2\rangle$ level, allowing discrimination between $S_{1/2}$ sublevels in few-ms detection durations without the need for a frequency-stabilized 729 nm `shelving' laser. We observe slow population of the `dark' $|S_{1/2},+1/2\rangle$ state with a time constant of $11.1(8)~\text{ms}$ with our laser beam parameters, which we attribute to off-resonant $S_{1/2 }\rightarrow P_{3/2}$ and $D \rightarrow P_{3/2}$ transitions. This depumping effect would be reduced in a trap with a larger magnetic field.

We apply a weak 397 nm laser beam as shown in Fig.~\ref{fig:CaBe}(a) to efficiently prepare the $|S_{1/2},-1/2\rangle$ state before microwave interrogation and for `bright' state repumping during continuous Doppler cooling. We find that the steady-state 393-nm fluorescence for this cooling method is within the same order of magnitude as the more typical high-field \Ca~laser cooling technique employing two 397~nm laser beams~\cite{koo_doppler_2004,mcmahon_doppler-cooled_2020}. 

Initial state preparation and detection of the co-trapped \Be~requires a single axially-oriented 313~nm laser beam tuned near the $|S_{1/2}, m_J=+1/2, m_I=+3/2\rangle \rightarrow |P_{3/2}, +3/2, +3/2\rangle$ electric dipole transition [see Fig.~\ref{fig:CaBe}(b)]. Here, $m_I$ is the projection of the \Be~nuclear spin ($I=3/2$) on the magnetic field axis. We produce the 313~nm light via a commercial laser system that relies on fourth harmonic generation of an amplified external cavity diode laser operating at $1252$~nm. Initial preparation of the $|S_{1/2},+1/2,+3/2\rangle$ `stretched state' after \Be~loading occurs over $\sim1~\text{s}$ due to off-resonant optical pumping from the 313 nm laser beam~\cite{itano_perpendicular_1988}.

All Doppler cooling and detection laser beams (313~nm, 393~nm, 397~nm, 850~nm, and 854~nm) are linearly polarized and are applied along both the axial and radial trap directions. To guarantee co-alignment of all laser beams, they are combined into a single beamline using a combination of dichroic mirrors and polarizing beamsplitter cubes. The combined beams are then split with a 50:50 beamsplitter; one half is directed along the trap $z$-axis while the other is coupled into a photonic crystal fiber (PCF) using a reflective collimator. We do not use a hydrogen-loaded PCF~\cite{colombe_single-mode_2014} for this work, because the 313~nm laser power is relatively low ($<100~\mu\text{W}$). The output of the PCF is directed to a parabolic reflector with a 15 cm focal length for co-focusing of all laser wavelengths in the radial direction. We measure a waist of $30~\mu\text{m}$ for the 397~nm laser beam at the ion crystal location. 

\section{\Ca~Spectroscopy}

Trapped \Ca~is useful for characterizing the Penning trap magnetic field environment due to the high sensitivity ($\sim28.025~\text{MHz/mT}$) of its $|S_{1/2},+1/2\rangle\rightarrow |S_{1/2},-1/2\rangle$ electron-spin resonance (ESR) transition~\cite{tommaseo_gj-factor_2003}. Also, in contrast to Be$^+$ and Mg$^+$, Ca$^+$ does not exhibit long-term loss from chemical reactions with residual H$_2$ molecules~\cite{sawyer_reversing_2015,condoluci_reassigning_2017}. We regularly maintain a \Ca~crystal for weeks. Undesired ion loss occurs when a Doppler cooling laser abruptly shifts to a blue-detuned longitudinal mode. 

\begin{figure}
	\includegraphics[scale=0.33]{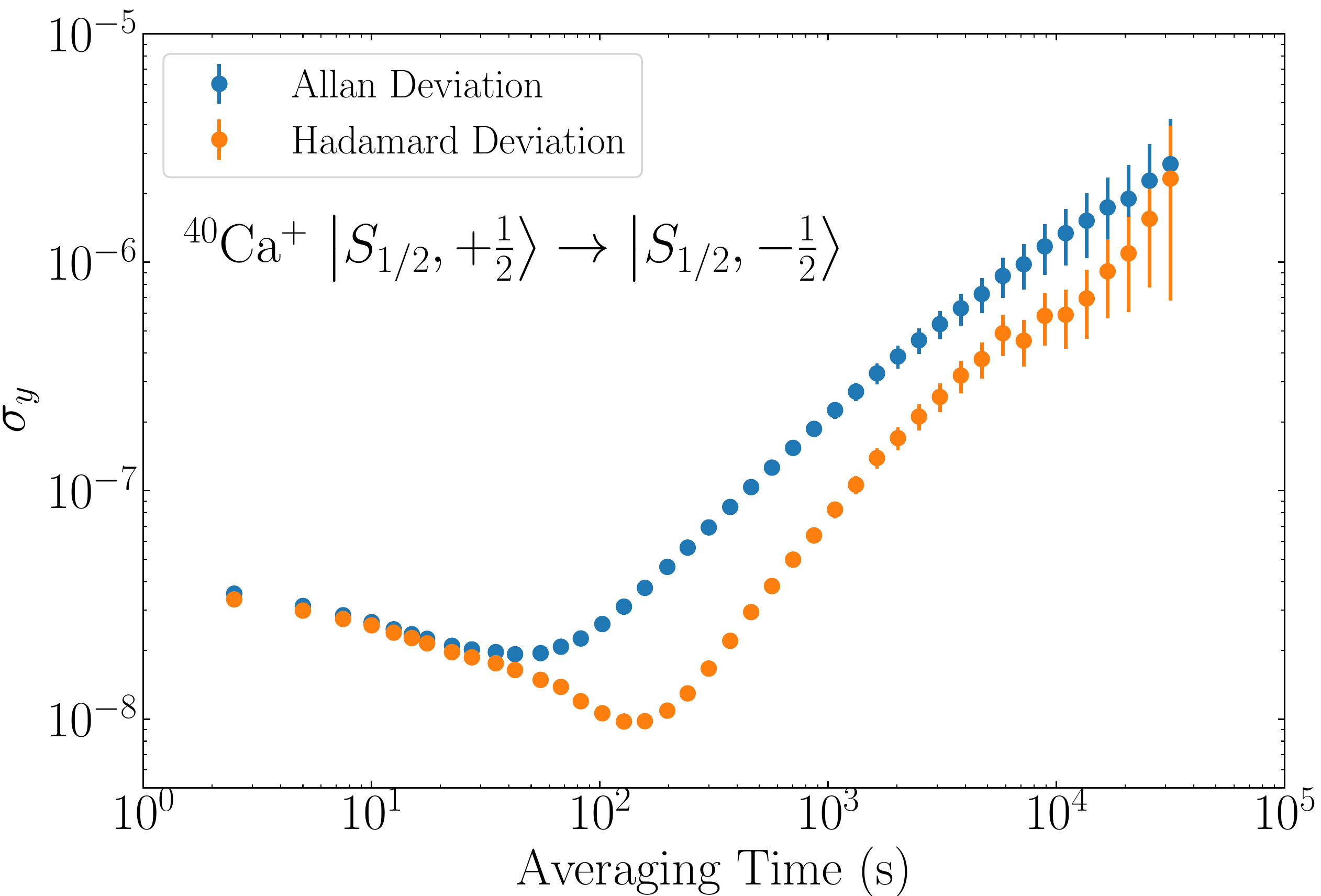}
	\caption{\label{fig:adev} Allan (blue points) and Hadamard (orange points) deviation of fractional frequency fluctuations ($\sigma_y$) versus averaging time computed from repeated measurements of the \Ca~electron spin resonance frequency near $18.3~\text{GHz}$. The fractional Allan deviation reaches a minimum of $19.2(3)$~ppb at $43$~s of averaging. The minimum fractional Hadamard deviation of $9.8(3)$~ppb is measured at $120~\text{s}-160~\text{s}$, corresponding to a drift-removed magnetic field instability of $6.4(2)$~nT.}
\end{figure}

We have quantified the instability of the \Ca~ESR frequency near $18.3~\text{GHz}$ in our PMPT over a 48 hour period. For these measurements, we apply $\sim 3$-W pulses of $18.3~\text{GHz}$ radiation via an external horn antenna directed orthogonally to the $z$ axis of the trap. The radial aperture of our Penning trap electrodes is 2~mm in the horizontal direction (see Fig.~\ref{fig:Trap}), which suppresses propagation of the $\sim1.6$-cm-wavelength microwave radiation by $>30~\text{dB}$. We observe Rabi lineshapes with a full width at half maximum (FWHM) of $\sim10~\text{kHz}$ using microwave pulse durations of $>1$~ms. We repeatedly sample the Rabi lineshape by measuring at three frequencies: two points spaced by the FWHM and a third point in between. The ESR central value is updated every $2.5~\text{s}$ based on the three-point measurement, which allows for characterization of long-term fluctuations of the PMPT magnetic field (see Fig.~\ref{fig:adev}).

In Fig.~\ref{fig:adev}, both the Allan deviation (ADEV) and Hadamard deviation (HDEV)~\cite{riley_handbook_1994} of \Ca~fractional ESR frequency fluctuations ($\sigma_y$) decrease for the first $\sim30~\text{s}$ of averaging, and the HDEV reaches a minimum of $9.8(3)~\text{ppb}$ in the range of $120~\text{s}-160~\text{s}$. Converting to absolute magnetic field, this HDEV implies a drift-removed minimum instability of $6.4(2)~\text{nT}$. Our apparatus does not currently include magnetic shielding, therefore our measured instabilities are the result of magnet remanence variation as well as ambient laboratory magnetic field fluctuations. The NdFeB remanence exhibits a fractional temperature sensitivity of $-1.2\times10^{-3} / \text{K}$, thus the minimum ADEV of $19.2(3)~\text{ppb}$ bounds the magnet temperature deviation to $<16~\mu\text{K}$ over $43~\text{s}$. This ADEV implies that active feedback (e.g. with an external magnet coil) via interleaved \Ca~co-magnetometry during \Be~interrogation would permit field stabilization at the level of $\sim20$~ppb; conferring a fractional frequency instability below $10^{-17}$ for \Be~at the MFI points (i.e. 0.6774~T or 0.8194~T).

\section{\Be~Spectroscopy}

Magnetic field shifts of the \Be~ground state hyperfine sublevels are well-described by the Breit-Rabi formula~\cite{shiga_diamagnetic_2011}. Excluding the stretched $[F=I+\frac{1}{2}$, $m_F=\pm(I+\frac{1}{2})]$ hyperfine states that remain uncoupled in an external magnetic field, we write the frequency shift ($\Delta f)$ of the remaining $S_{1/2}$ hyperfine sublevels as 
\begin{eqnarray}
    &\Delta& f(m_F, X) = \nonumber \\ &A& \left( -\frac{1}{4} + 2\frac{\gamma}{1-\gamma}m_F X
    \pm \sqrt{1 + m_F X + X^2} \right),
\end{eqnarray}
where $m_F=m_I + m_J$ is the projection of total angular momentum on the magnetic field axis, $A$ is the \Be~ground state hyperfine constant, $\gamma=g_I'/g_J$ is the \Be~g-factor ratio of nuclear and electron spins, $X\equiv\mu_b B (g_J - g_I')/[(I+1/2)hA]$, $\mu_b$ is the Bohr magneton, and $h$ is Planck's constant. 

Using the \Be~hyperfine constants and nuclear gyromagnetic ratio reported in Ref.~\cite{shiga_diamagnetic_2011}, we calculate first- and second-order sensitivities of $-369.5~\text{Hz/mT}$ and $16.5~\text{Hz/mT}^2$, respectively, between the $|S_{1/2}, -1/2, +3/2\rangle$ and $|S_{1/2}, -1/2, +1/2\rangle$ states at our nominal trap field of 0.6540~T. For comparison, at 0.677396~T the first-order sensitivity of this transition approaches zero and the second-order term reduces slightly to $15.1~\text{Hz/mT}^2$. While the NdFeB permanent magnet dimensions were chosen to achieve a uniform magnetic field of 0.6774~T with the specified N52 remanence, the actual measured remanence is $\sim5\%$ lower. A trap field of 0.6774~T is still achievable via adjustment of the axial spacing of the ring magnets~\cite{mcmahon_doppler-cooled_2020}. However, the increased magnitude would come at the expense of reduced trap magnetic field uniformity. 

For the \Be~clock operation of Ref.~\cite{bollinger_303-mhz_1991}, the authors used a combination of resonant microwave and radiofrequency fields to transfer population from the $|S_{1/2},+1/2,+3/2\rangle$ Doppler cooling state to their clock transition manifold. Given the high degree of microwave shielding at $\sim18~\text{GHz}$ that is induced by our trap electrode geometry, we have chosen to instead use incoherent optical pumping to populate our MFI manifold, which consists of the $|S_{1/2},-1/2,+3/2\rangle$ and $|S_{1/2},-1/2,+1/2\rangle$ levels. This optical pumping scheme is insensitive to slow drifts of the trap magnetic field at the level of $\sim0.1~\text{mT}$. We produce the 313~nm optical pumping tone shown in Fig.~\ref{fig:CaBe}(b) by passing the detection laser beam through a free-space EOM with an adjustable resonance frequency of $6.07\pm0.04~\text{GHz}$. The EOM resonance is tuned so that the $-2$ sideband excites to the $|P_{3/2}, -1/2, +3/2\rangle$ state, which decays via electric dipole radiation primarily to the unexcited $|S_{1/2},-1/2,+3/2\rangle$ state. Using a rate equation model, we predict an optical transfer efficiency to the state of interest of $>99\%$. We typically observe $\sim70\%$ transfer efficiency from $|S_{1/2},+1/2,+3/2\rangle$ to $|S_{1/2},-1/2,+3/2\rangle$, which we attribute to uncorrected AC Stark shifts from the optical carrier and $\{\pm 1, +2\}$ EOM sidebands as well as imperfect overlap of the radially-oriented laser beam with the \Be~crystal.

\begin{figure}
	\includegraphics[scale=0.31]{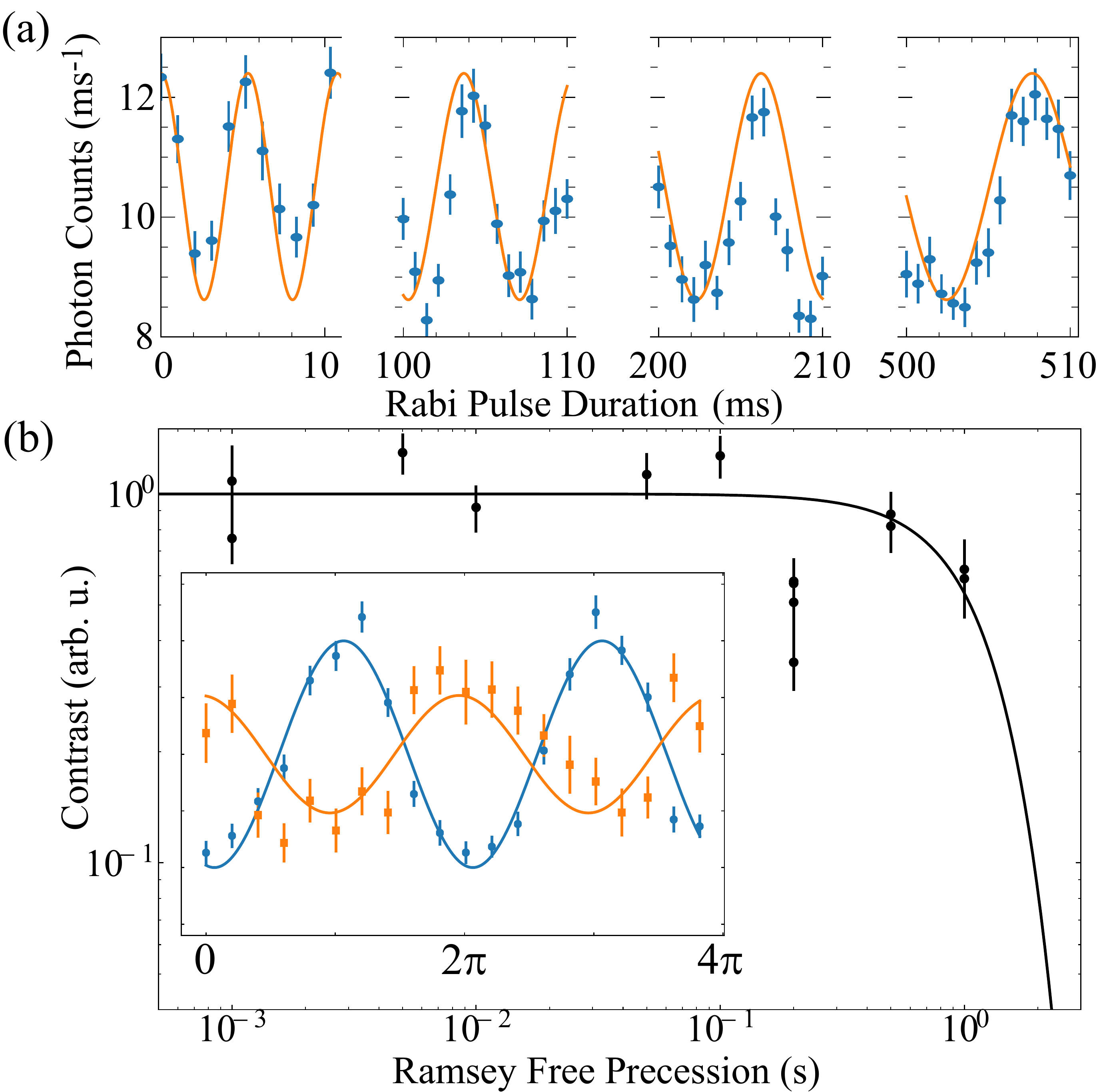}
	\caption{\label{fig:clockRamsey} (a) Resonant Rabi interrogation of the magnetic-field-insensitive radiofrequency transition in \Be over $>500~\text{ms}$ (blue points with error bars) and fit (solid orange line). (b) Measured Ramsey fringe contrast versus free precession time at $321.172~691~\text{MHz}$, which is consistent with a trap magnetic field of $0.653681~\text{T}$. Black points with error bars are the measured contrast, and a fit to a Gaussian decay model (solid black line) yields a decay time of \coherence. Based on the measurements of Fig.~\ref{fig:adev}, we attribute the observed contrast decay to uncorrected magnetic field drift during data collection at extended free precession periods. (Inset) Example \Be~fluorescence vs. Ramsey analysis phase data with associated sinusoidal fits for 1~ms (round blue points) and 1~s (square orange points) free precession periods.}
\end{figure}

After optically pumping the \Be~to $|S_{1/2},-1/2,+3/2\rangle$, we then transport the combined \Be--\Ca crystal radially by $\sim1~\text{mm}$ from the laser cooling and detection position to a trap region with minimum magnetic field curvature~\footnote{This position of minimum magnetic field curvature suffered from a high background scatter rate from the 313~nm detection laser beam. We chose to use ion transport to reach this location in lieu of precisely repositioning the permanent magnet array.}. We perform ion transport by modifying individual trap electrode voltages using programmable digital-to-analog converters. The optimal trap location is determined by mapping the local trap magnetic field in three dimensions using a small, $\sim20$-ion crystal of \Ca~via ESR measurements. All trap electrodes are filtered using two-stage RC low-pass filters with a corner frequency of 15.5~kHz, which facilitates adiabatic transport of the ion crystal at timescales of $\sim100~\mu\text{s}$.       

We interrogate the 321-MHz MFI transition using a RF coil oriented radially and placed outside a vacuum chamber viewport. The RF coil consists of a single loop of copper wire with an inductance of $\sim60~\text{nH}$ placed in parallel with a variable capacitor tuned near 4~pF. This resonant LC circuit is AC-coupled to a high-power RF amplifier via a 1-pF capacitor placed in series. Figure~\ref{fig:clockRamsey}(a) shows Rabi flops of the MFI transition with a fit $\pi$-pulse time ($t_{\pi}$) of 2.6(1)~ms. We observe coherent Rabi nutation for $>0.5~\text{s}$ with no loss of contrast, however the data does show a monotonic reduction in the resonant Rabi rate for increasing interaction times. We include this effect in the fit (orange line) of Fig.~\ref{fig:clockRamsey}(a) via a phenomonological power-law decay of the Rabi rate in time. This slow decay of RF power is likely the result of heating in the high-power amplifier and resonant coil.

Figure~\ref{fig:clockRamsey}(b) shows Ramsey contrast versus free precession time measurements. As in the Rabi flop experiments of Fig.~\ref{fig:clockRamsey}(a), we begin by optically pumping to the $|S_{1/2},-1/2,+3/2\rangle$ state. We then pulse the resonant RF for $t_{\pi}/2$ (i.e. $\pi/2$ pulse) to create an equal superposition of the $|S_{1/2},-1/2,+3/2\rangle$ and $|S_{1/2},-1/2,+1/2\rangle$ states. The spins then precess for a variable time before a second RF $\pi/2$ `analysis' pulse is applied with a variable relative phase from the first in the range $[0, 4\pi]$. Finally, we repump the population of $|S_{1/2},-1/2,+3/2\rangle$ back to the bright state over a $10~\text{ms}$ duration before detecting for $2.5~\text{ms}$. We observe a repumping time constant of $4.6(9)~\text{ms}$ for transitions from $|S_{1/2},-1/2,+3/2\rangle$ to $|S_{1/2},+1/2,+3/2\rangle$ when applying the 313~nm detection laser beam. In contrast, we measure a time constant of $190(40)~\text{ms}$ for repumping from $|S_{1/2},-1/2,+1/2\rangle$ back to the bright $|S_{1/2},+1/2,+3/2\rangle$ state. This $>40\times$ ratio of repumping times facilitates MFI state discrimination using only the detection laser beam.

The resulting photon counts vs. analysis pulse phase are fit to a sinusoid [see inset of Fig.~\ref{fig:clockRamsey}(b)] whose amplitude is plotted in Fig.~\ref{fig:clockRamsey}(b) at various Ramsey free precession times. To compare multiple data sets taken over different days with varying numbers of \Be, each contrast value shown is computed relative to that measured at $10~\mu\text{s}$ of free precession. The inset of Fig.~\ref{fig:clockRamsey}(b) shows sample Ramsey phase scans for $1~\text{ms}$ and $1~\text{s}$ of free precession as blue points and orange squares, respectively. Assuming a Gaussian contrast decay model, we obtain a $1/e$ decay time constant of \coherence. For the the longest free precession durations, we observe that subsequent scans are phase-shifted due to the slow drift of the trap magnetic field presented in Fig.~\ref{fig:adev}. With the addition of active magnetic field stabilization to $20~\text{ppb}$ via repeated interrogation of the \Ca~ESR transition, we expect $>200~\text{s}$ of \Be~spin coherence in this system in spite of the $\sim-370~\text{Hz/mT}$ residual linear magnetic field sensitivity. While not required for this demonstration, we expect that extended ($\gg 1~\text{s}$) Ramsey free precession durations will necessitate continuous sympathetic cooling as in Ref.~\cite{bollinger_303-mhz_1991}.

\section{Conclusions}

We have performed the first spectroscopic characterization of the magnetic field instability of a PMPT. Demonstrating co-trapping of \Be--\Ca~crystals, we present precision spectroscopy of both species, achieving a $>1$-s quantum coherence of the \Be~MFI transition. Active magnetic field stabilization at our realized fractional instability of $19.2(3)~\text{ppb}$ would permit \Be~clock operation with a fractional precision as small as $10^{-17}$ at a trap magnetic field of $0.6774~\text{T}$ or $0.8194~\text{T}$. Systematic fractional frequency offsets of the 321-MHz clock transition would include AC Stark shifts from the \Ca~Doppler cooling laser beams of $<10^{-15}$ and, more fundamentally, a minimum \Be~second-order Doppler shift of $\sim-1\times10^{-15}$ for $100$ ions with our trap parameters~\cite{tan_minimizing_1995}. Intensity stabilization of the Doppler laser beams along with rotating wall control~\cite{huang_phase-locked_1998} of the ion crystal aspect ratio allows straightforward stabilization of these systematic shifts to $<10^{-16}$. We note that the number of clock ions should ultimately be increased to reduce quantum projection noise~\footnote{The quantum projection noise scales as $N^{-1/2}$, while the minimum second-order-Doppler shift scales as $N^{2/3}$, where $N$ is the number of clock ions.}. Future experiments could elucidate the previously-observed \Be~clock accuracy limit of $1\times10^{-13}$, which Bollinger and co-authors attributed to collisions with background CH$_4$ molecules~\cite{bollinger_303-mhz_1991}. Furthermore, an alternative clock species such as $^{43}\text{Ca}^+$ ($I=7/2$) would mitigate reactive ion loss~\cite{condoluci_reassigning_2017} and provide clock transitions at higher trap magnetic field.

Recent work with \Be~in a superconducting-magnet-based Penning trap highlights the unique sensitivity of high-field ESR spectroscopy to magnetic field fluctuations~\cite{britton_vibration-induced_2016}. Our measured PMPT fractional field instability of $<20$~ppb is consistent with previous low-field (0.37~mT) measurements utilizing permanent magnets~\cite{ruster_long-lived_2016}. Improved measurement sensitivities of $<1~\text{ppb}$ would facilitate characterization of the spectrum of intrinsic field fluctuations within rare-earth permanent magnet materials such as NdFeB or SmCo~\cite{schloemann_fluctuations_1984}. Such a system might serve as a magnetic analogue to the trapped-ion electric field sensors used for characterizing metals~\cite{hite_100-fold_2012,mcconnell_reduction_2015,sedlacek_evidence_2018,mckay_measurement_2021} and dielectrics~\cite{teller_heating_2021}. 

\begin{acknowledgments}
We thank John J. Bollinger, Nicholas D. Guise, and Christopher Shappert for helpful discussions as well as Kenton R. Brown, Robert Wyllie, and Creston Herold for comments on the manuscript. We also acknowledge Curtis Volin for designing the imaging objective. We thank Dietrich G. Leibfried and Jonas Keller for providing beryllium ovens for these experiments. This work is funded by Office of Naval Research grant N00014-17-1-2408. 
\end{acknowledgments}

\bibliography{PenningRefsBJM} % Produces the bibliography via BibTeX.

\end{document}